\documentclass[aps,pre,twocolumn,superscriptaddress]{revtex4}
\usepackage{graphicx,amsmath} 
\usepackage[amssymb]{SIunits} 
\usepackage{color}
\usepackage[dvipsnames]{xcolor}
\usepackage{multirow}
\usepackage{placeins}
\begin{document}
\title{Mixing induced by a bubble swarm rising through incident turbulence}
\author{Elise~Alm\'eras}\thanks{elise.almeras@ensiacet.fr}
\affiliation{Laboratoire de G\'enie Chimique, UMR 5503, CNRS-INP-UPS, 31106 Toulouse, France}
\affiliation{Physics of Fluids Group, Faculty of Science and Technology, J. M. Burgers Centre for Fluid Dynamics, University of Twente, P.O. Box 217, 7500 AE Enschede, The Netherlands}

\author{Varghese Mathai}
\affiliation{School of Engineering, Brown University, 02912 Providence, RI, USA}
\affiliation{Physics of Fluids Group, Faculty of Science and Technology, J. M. Burgers Centre for Fluid Dynamics, University of Twente, P.O. Box 217, 7500 AE Enschede, The Netherlands}

\author{Chao Sun}
\affiliation{Center for Combustion Energy and Department of Thermal Engineering, Tsinghua University,  100084 Beijing, China}
\affiliation{Physics of Fluids Group, Faculty of Science and Technology, J. M. Burgers Centre for Fluid Dynamics, University of Twente, P.O. Box 217, 7500 AE Enschede, The Netherlands}

\author{Detlef Lohse}
\affiliation{Physics of Fluids Group, Faculty of Science and Technology, J. M. Burgers Centre for Fluid Dynamics, University of Twente, P.O. Box 217, 7500 AE Enschede, The Netherlands}
\affiliation{Max Planck Institute for Dynamics and Self-Organization, 37077 G\"ottingen, Germany}

\date{\today}

\begin{abstract} 
This work describes an experimental investigation on the mixing induced by a swarm of high Reynolds number air bubbles rising through a nearly homogeneous and isotropic turbulent flow. The gas volume fraction $\alpha$ and the velocity fluctuations $u'_0$ of the carrier flow before bubble injection are varied, respectively, in the ranges $0\leq \alpha \leq 0.93\%$ and 2.3 cm/s $\leq u'_0 \leq 5.5$ cm/s, resulting in a variation of the bubblance parameter $b$ in the range [0, 1.3] ($b = \frac{V_r^2 \alpha}{{u'}_0^2}$, where $V_r$ is the relative rising velocity). Mixing in the horizontal direction can be modelled as a diffusive process, with an effective diffusivity $D_{xx}$. Two different diffusion regimes are observed experimentally, depending on the turbulence intensity. At low turbulence levels, $D_{xx}$ increases with gas volume fraction $\alpha$, while at high turbulence levels the enhancement in $D_{xx}$ is negligible. When normalizing by the time scale of successive bubble passage, the effective diffusivity can be modelled as a sole function of the gas volume fraction $\alpha^* \equiv \alpha/\alpha_c$, where $\alpha_c$ is a theoretically estimated critical gas volume fraction. The present explorative study provides insights into modeling the mixing induced by high Reynolds number bubbles in turbulent flows.
\end{abstract}

\maketitle

\section{Introduction}

Air bubble-laden liquid flows are commonly found in many natural, chemical and biological processes. Just as any freely rising body in a liquid, the rising air bubbles in these flows can move along oscillatory paths through the surrounding liquid~\citep{Magnaudet_annual_review,magnaudet2007wake,zenit2008path,ern2012wake,mathai2015wake}.
These wake-induced oscillations are known to generate agitation in the liquid phase, and are often used as an efficient way of mixing different chemical and biological components avoiding the need for mechanical mixing devices~\citep{DARMANA_ces,balachandar2010turbulent,risso2018agitation}. 

While there is general consensus that high Reynolds number bubbles enhance the liquid agitation~\citep{aybers1969motion,biesheuvel1984two,bunner2002dynamics,rensen2005effect,cartellier2009induced,mercado2010bubble,roghair2011energy,roig2012dynamics,ziegenhein2017towards}, predicting the mixing due to the bubbles remains a challenging task. This is because in these flows, the bubble induced mixing results not only from the agitation by the bubbles, but also are due to the transport by the large scale recirculations, the bubble wakes, and the shear. A few studies have been conducted to disentangle the different mixing mechanisms in a bubble column, and to understand particularly the mixing induced by bubbles~\citep{mareuge1995,Expfluidbouche2013,thesealmeras,JFMloisy2018}. These have shown that mixing induced by bubbles results from two contributions:~\textit{(i)}~the mixing by capture in the bubble wakes and \textit{(ii)}~the mixing by the agitation resulting from the wake-wake interactions. Mixing by dye captured in the bubble wakes is an intermittent and convective process, and is dominant in a 2D homogeneous bubbly flow rising in a Hele-Shaw cell since no interaction between the wakes can develop due to confinement~\citep{Expfluidbouche2013,almeras2016time,PRFalmeras2018}. On the contrary, mixing induced by the agitation resulting from the wake-wake interactions is the leading mixing mechanism in three-dimensional bubble columns. This is a diffusive process and can thus be described by an effective diffusivity \citep{mareuge1995,JFMalmeras2015,gvozdic2018experimental}. Following Taylor theory, Alm\'eras {\it et al.}\cite{JFMalmeras2015} proposed to model the diffusion coefficient, as $D_{ii}={u'}^{2}_{i} \ T$, where ${u'}^{2}_{i}$ is the variance of the velocity fluctuations of the liquid phase, and $T$ is a characteristic time scale of the turbulent diffusion that requires to be reinterpreted for two-phases flows. Namely, at low gas volume fractions, this characteristic time scale is given by the Eulerian integral length scale $\lambda$ of the bubble-induced turbulence, $T=\frac{\lambda}{u'}$, while it is the time between two bubbles passages $T_{2b}$ at larger gas volume fractions~\citep{JFMalmeras2015}.

\begin{figure*} [!htbp] \centering
	\centerline{\includegraphics[scale=2.7]{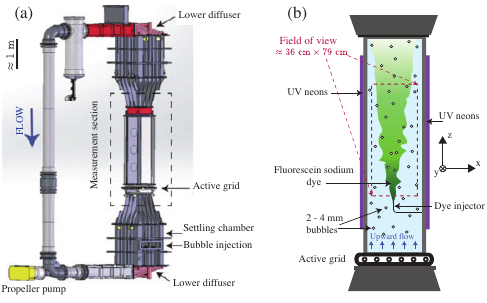}}
	\caption{(a) Schematic of the Twente-Water-Tunnel where the mixing experiments were performed. (b) A zoom-in view shows the measurement section with the active grid below it, and an upward bubbly turbulent flow. UV neons are used to illuminate the injected fluorescent dye. Two sCMOS cameras (not shown here) are placed in front of the measurement section, yielding a total field of view of 36 cm $\times$ 79 cm, as shown in the schematic.}
	\label{fig:exp_setup}
\end{figure*}

As evident from the above discussion, there exists a wealth of experimental and numerical investigations on homogeneous bubble swarms rising within an otherwise  still liquid. However, in most natural and industrial settings~(upper ocean mixing layer, industrial bubble columns, etc), the bubble swarms rise within surrounding turbulence. Several studies were devoted to the hydrodynamics properties of such turbulent bubbly flows~\citep{jfmlance1991,rensen2005effect,prakash2016energy,JFMalmeras2017,mathai2018dispersion}. These have addressed the problem from both Eulerian and Lagrangian viewpoints. Alm\'eras \textit{et al.}\cite{JFMalmeras2017} showed that the liquid agitation in such flows are the outcome of a complex interaction between the bubble swarm and external turbulence at the large scales. Recently, in a Lagrangian investigation of 2 mm bubbles in homogeneous turbulence, Mathai {\it et al.} \cite{mathai2018dispersion} reported that bubbles contribute to a fast short term spreading, but a slower large scale spreading. The relevant competing time scale for the large scale spreading was found to be the crossing time of the bubbles as compared to the integral time scale of the background turbulence~\citep{parishani2015effects,mathai2016microbubbles}. 

While the above studies provided insight into the bubble motions, spreading, and the resulting liquid agitation in turbulent flows, they did not address specifically the mixing induced by high Reynolds number bubbles in such tubulent flows. In the present work, we address this question by analysing the mixing of a passive scalar~(a high Schmidt number dye) in a homogenous bubble swarm rising within a nearly homogeneous and isotropic turbulent flow.

The paper is organised as follows. In section~\ref{sec:exp}, the experimental setup and the measurement techniques used for the mixing experiments are presented, while the hydrodynamics properties of the turbulent bubbly flow are discussed in section~\ref{sec:hydro}. The mixing properties of the turbulent bubbly flow are described in section~\ref{sec:mixing}. In section~\ref{sec:discussion}, the mixing model is discussed. Finally, concluding remarks are presented in section \ref{sec:conc}.

\section{Experimental set-up and instrumentation}
\label{sec:exp}

\begin{table}[!b]
	\centering
	\begin{tabular}{p{0.1\linewidth}p{0.1\linewidth}p{0.1\linewidth}p{0.10\linewidth}p{0.1\linewidth}p{0.1\linewidth}}
		$U$ & $u'_0$ & Re$_\lambda$  &   $\tau_\eta$ & $T_L$ \\
		\hline
		m$/$s & cm$/$s & - &  s &  s \\
		\hline
		0.27  & 2.3 & 177 & 0.083 & 3.6 \\
		0.27  & 3.1 & 242 &  0.066 & 3.9 \\
		0.27  & 3.3 & 262 &  0.061 & 3.9\\
		0.26  & 3.4 & 265 &  0.059 & 3.8\\
		0.46  & 3.5 & 216 &  0.044 & 2.3 \\
		0.46  & 4.6 & 315 &  0.037 & 2.9 \\
		0.46  & 5.1 & 342 &  0.034 & 2.8 \\
		0.46  & 5.5 & 361 &  0.030 & 2.6 \\
	\end{tabular} 
	\caption{Summary of the flow parameters for single-phase flow in the Twente-Water-Tunnel facility. Here $U$ is the mean flow velocity, $u_0'$ is the standard deviation of velocity fluctuations, Re$_{\lambda}$ is the Taylor-Reynolds number, and $\tau_\eta$ and $T_L$ are the dissipative and integral time scales of the turbulent flow, respectively. }
	\label{tab:singlephase}
\end{table}

Mixing experiments are performed in the Twente Water tunnel (figure~\ref{fig:exp_setup}(a)), which has a vertically long measurement section of 2 m height, 450$\times$450 mm$^2$ in cross-section, and made of three glass walls allowing optical access. An active-grid is positioned at the bottom of the measurement section, and an upward flow generates nearly homogeneous and isotropic turbulence at a distance roughly 1~m above the grid location. The level of the external turbulence is varied by adjusting both the liquid mean flow speed and the rotation speed of the active grid. In the present study, eight levels of turbulence are investigated. These have been characterised by looking at the statistical properties of the liquid phase in the absence of bubbles (see Ref.~\cite{JFMalmeras2017} for more details). Table~\ref{tab:singlephase} summarises the flow characterisation of the eight single phase turbulence cases investigated in the present study. The standard deviation of the velocity fluctuations produced by external turbulence in the absence of bubbles ranges from $u'_0=2.3$ cm/s to $u'_0=5.5$~cm/s, which was achieved by means of two mean flows ($U=0.27$ m/s and $U=0.46$ m/s) and four different grid rotation speeds. Note that $T_L$ is the Eulerian integral time scale of the flow. We expect this to be comparable to the Lagrangian integral time scale, which is relevant for turbulent mixing~\citep{taylor}.

A homogeneous bubble swarm is produced using 621 capillary tubes of inner diameter 0.12 mm positioned on 9 islands, which are located below the active-grid~\citep{JFMalmeras2017}. By varying the gas flow rate, the gas volume fraction $\alpha$ can be adjusted from 0 to 0.93$\%$. The air bubble injection setup generates bubbles with a diameter $d$ ranging from 2 to 3.3 mm, and an aspect ratio $\chi$ varing from 1.6 to 2.0, depending on the turbulence level and the gas volume fraction. The bubble relative velocity $V_r$ turns out to decrease with increasing gas volume fraction, irrespective of the turbulence level, from 0.39 cm$/$s to 0.27 cm$/$s. More details of the gas phase properties for identical operating conditions are given in Ref.~\cite{JFMalmeras2017}.

The setup for the mixing experiments is shown in figure~\ref{fig:exp_setup}(b). The experiment involves injecting continuously a fluorescent dye (Fluorescein sodium) within the flow~\citep{almeras2016time,mathai2018flutter} by means of a dye injector of diameter 2 mm located in the middle of the cross-section and 60 cm downstream of the active-grid. The dye has a high Schmidt number ($S_c \equiv \frac{\nu}{D} \approx 2500$) in water, and hence, can be considered passive scalar in the flow. The dye injection is performed for a duration of 50 s at a flow rate matching the mean liquid velocity in the tunnel. In this way, the injection of the dye does not cause a strong jet, which allows us to study the dye dispersion effects due to the bubbly turbulent flow alone. The concentration of fluorescein sodium at the injection location is $c_0 \approx 10^{-2}$ mol/L, and has been chosen in such way that the mixing of the dye could be visualised over approximately 1 m. {\color{black}The field of view lies in the $zx$ plane, with $z$ the vertical (streamwise) direction, $x$ one of the horizontal (transverse) directions, and the corresponding liquid velocity components $u$ and $v$, respectively. The depth direction is given by $y$ (see figure~\ref{fig:exp_setup}(b)).} Due to the high concentration of dye, absorption of the light by the dye occurs close to the injector. This effect is however negligible far enough from the dye injector. The minimum distance at which the measurement is valid is determined for each operating condition by calculating the total mass along the height (see section \ref{sec:mixing}). This also ensured that any disturbances from the injection had dissipated at the location of the measurements.

The instantaneous spatial dye distribution has been measured by means of fluorescence induced by UV neon tubes, as already introduced by \cite{JFMalmeras2015} for measuring the temporal evolution of the dye concentration distribution in a homogeneous bubbly flow. To this end, eight UV neon tubes are placed on opposite sides of the cross-section of measurement (figure \ref{fig:exp_setup} (b)). This light configuration allows to have a homogeneous illumination over the full cross-section measurement spanning a total height of 120 cm, with a tolerance range of 12$\%$. Two synchronised cameras (Imager sCMOS from Lavision) equiped with 105 mm macro lenses and optical band-pass filters ($450 - 650$ nm) are aligned vertically. Fluorescent light emitted by the dye is recorded over a total field of view of 79.0 cm height and 36.5 cm wide with a spatial resolution of 57 pixels/cm (figure \ref{fig:exp_setup}). The depth of view of the camera is maintained sufficiently large so that the magnification along the width is negligible. The integration time of the cameras is 10 ms, which results in 500 images in the 50 s of recording. A typical image recorded by this system is given in figure \ref{fig:concentration_profile}(a).  
For the provided specific conditions, the intensity of the fluoresced light emitted by the dye is proportional to the dye concentration \cite{crimaldi}. However, in a bubbly flow, two kinds of measurement perturbations occur due to the presence of bubbles \cite{JFMalmeras2015}. On one hand, reflections and refractions of the fluoresced light on the bubble interfaces outside the dye distribution produce a spreading of the fluoresced light viewed by the camera. On the other hand, occultations of the fluoresced light by the bubbles within the dye distribution result in an attenuation of the fluoresced light filmed by the camera. Therefore, an image processing method needs to be applied in order to measure the unperturbed concentration distribution. We adapt the image processing developed by \cite{JFMalmeras2015} to our case so that we are able to deal with the present liquid mean flow. The main adaptation consists in tracking the spatial distribution of the dye instead of its temporal evolution. In the present image processing, reflection and refraction effects are corrected by calculating at each height $z$ the minimum of gray levels in the vertical direction over the distance $[z-\frac{\Delta H}{2} : z+ \frac{\Delta H}{2}]$. The effect of the occultations is treated by calculating at each height $z$, the maximum of gray level in the vertical direction over the distance $[z-\frac{\Delta H}{2} : z+ \frac{\Delta H}{2}]$. The distance $\Delta H$ has been chosen to be larger than the typical size of the perturbations, which is of the order of the bubble diameter $d$ but small enough to keep a satisfactory spatial resolution of the measurement. We take $\Delta H=3.5$ mm. Validation of the image processing has been performed similar to the procedures followed in \cite{JFMalmeras2015}, i.e. by immersing a transparent plastic sphere of 6 cm diameter filled with a solution of fluorescein at a given concentration within the turbulent bubbly flow. This ensures a reliable measurement of instantaneous concentration distribution in the horizontal direction for $\alpha \leq 1\%$. Note that the instantaneous concentration profiles are shifted in order to center all the profiles on their barycenter before time averaging. In this way, only the spreading of the dye distribution around its global motion is measured.

\section{Hydrodynamic properties of the turbulent bubbly flow}
\label{sec:hydro}
\begin{figure*} [!htbp]
	\centering
	\includegraphics[scale=.55]{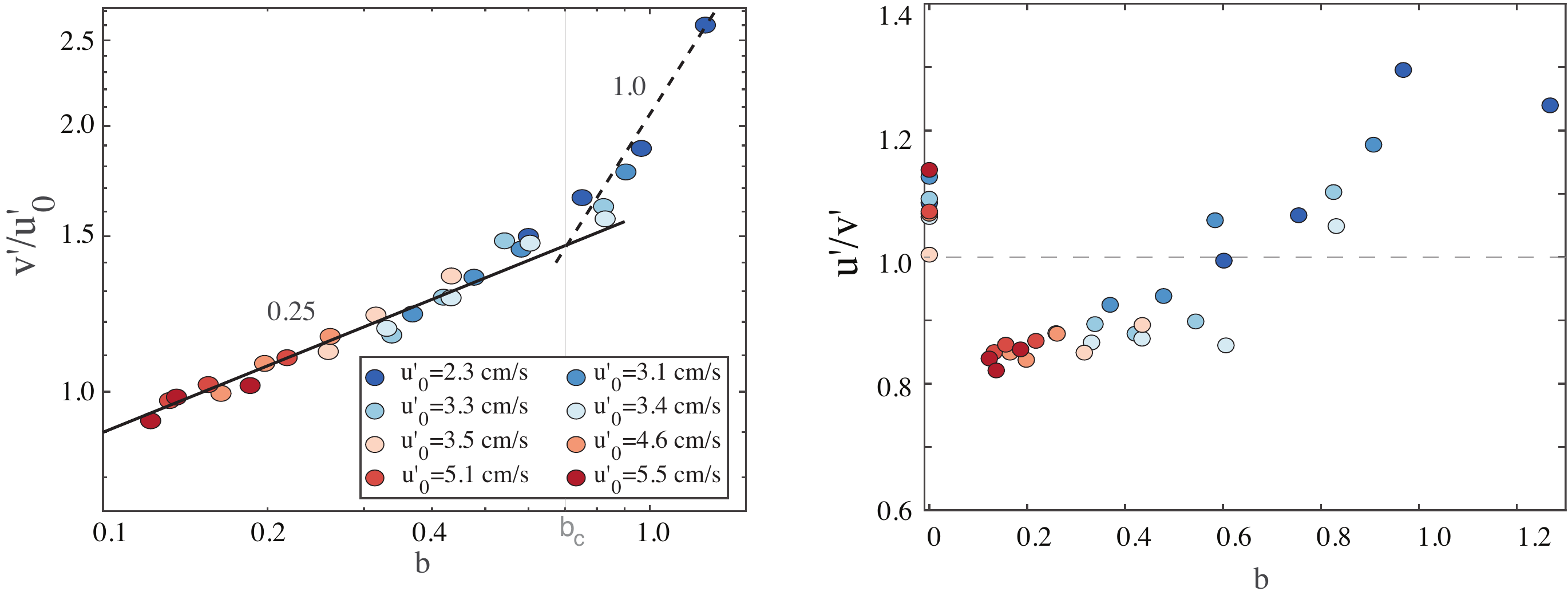}
	\caption{(a)Standard deviation of the velocity fluctuations $v'$ in the horizontal direction normalised by $u'_0$ as a function of the bubblance parameter $b$ on log-log scale. The error bars are $\leq 0.25 \ v'/u_0'$, i.~e. comparable to those reported by \cite{JFMalmeras2017}. Figure (b) shows the anisotropy of the vertical to horizontal velocity fluctuations as a function of $b$. The color code in both figures reflects the value of the incident turbulent fluctutations without bubbles~(See also table~\ref{tab:singlephase}).}
	\label{fig:vel_fluct}
\end{figure*}

Understanding the dispersion properties of a flow requires first a clear description of the hydrodynamic properties of the liquid phase. For the same set of turbulence levels and gas volume fractions, Alm\'eras {\it et al.}\cite{JFMalmeras2017} had performed an experimental characterisation of the liquid agitation. The main results of this study are recalled as: \\

The agitation of the liquid phase in the vertical direction is mainly controlled by the so-called bubblance parameter \citep{jfmlance1991,rensen2005effect,van2006turbulent},
\begin{equation}
b=\frac{V_r^2 \alpha}{{u'_0}^2}
\end{equation}  
which compares the ratio of the velocity fluctuations produced by the bubble swarm to the one produced by the external turbulence. Further, two regimes of liquid agitation were identified for the vertical velocity fluctuations, separated by $b \approx 0.7$. However, the horizontal liquid velocity fluctuations, which is crucial to model the horizontal mixing, were not characterized~\cite{JFMalmeras2017}. Here we have performed additional Laser Doppler Anemometry~(LDA) measurements in order to characterise the velocity fluctuations in the horizontal direction as well. Once normalized by the turbulent velocity fluctuations $u'_0$, the standard deviation of the velocity fluctuations $v'$ in the horizontal direction increases with $b$ (figure \ref{fig:vel_fluct}). Similar to the velocity fluctuations in the vertical direction (reported by Alm\'eras {\it et al.}\cite{JFMalmeras2017}), we can observe the presence of two regimes separated by $b\approx 0.7$. The anisotropy ratio of the velocity fluctuations is also characterised as a function of the bubblance parameter in the inset to figure \ref{fig:vel_fluct}. We see that it increases monotonically from 0.8 to 1.2 for $b$ ranging from 0.13 to 1.3, with almost isotropic fluctuations around $b \approx 0.7$. This increasing trend of the anisotropy ratio is consistent with literature, since the anisotropy ratio is known to approach 1.3 at $b \to \infty$, i.e. for pseudo-turbulence~\citep{JFMriboux}.  

It is important to stress that in the present turbulent bubbly flow, there is a clear separation of the times scales produced by the incident turbulence with the one generated by the bubble swarm. In fact, the external turbulence involves time scales ranging from the dissipative time scale $\tau_\eta$ to the integral time scale $T_L$ of the turbulent flow, governed by the Taylor-Reynolds number Re$_\lambda$~(see also Table~\ref{tab:singlephase}). In contrast, the bubble swarm affects the flow at shorter {\color{black} time scales $\tau_b \sim {d}/({C_d  V_r})$}, where $C_d$ is the drag coefficient of a bubble within the swarm \cite{JFMriboux}. We note that at all operating conditions of the present study, the bubble induced time scale is shorter than the dissipative time scale of the turbulence, \textit{id est} the Kolmogorv time scale.

\section{Mixing characterization in turbulent bubbly flow }
\label{sec:mixing}

\begin{figure*}
	\centering
	\includegraphics[scale=0.55]{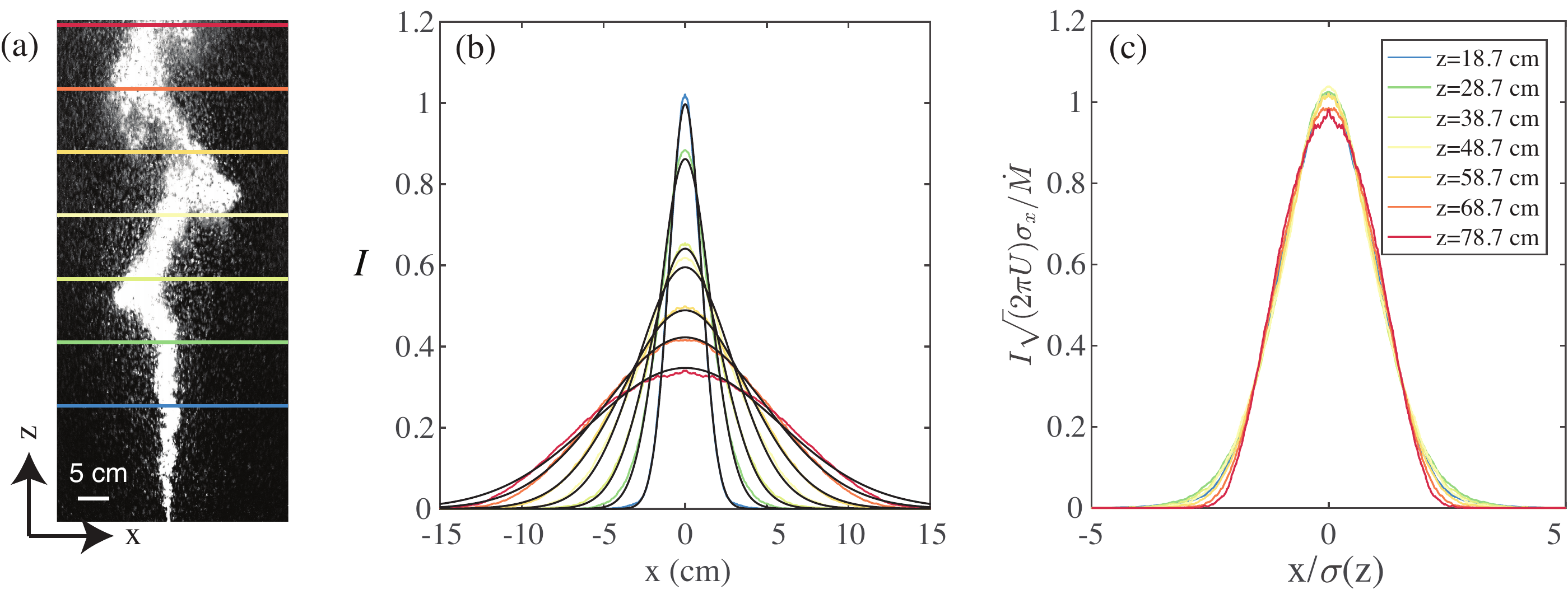}
	\caption{Dispersion of a fluorescent dye in a turbulent bubbly flow for $u'_0=2.3$ cm/s and $\alpha = 0.68\%$. (a) Instantaneous spatial distribution of the fluoresced light before image processing. (b) Experimental time-averaged concentration profiles at different heights (in colour) and Gaussian fits (in black). (c) Time-averaged concentration profiles normalized by the mass $\sqrt{2\pi U} \sigma_x(z) / \dot{M}$ as a function of the horizontal position normalized by the standard deviation $\sigma(z)$.}
	\label{fig:concentration_profile}
\end{figure*}

Figure \ref{fig:concentration_profile} presents the concentration profiles $I$ averaged over time at different heights $z$ for a turbulence intensity $u'_0=2.3$ cm/s and a gas volume fraction $\alpha=0.68\%$, corresponding to $b=0.94$. Due to mixing induced by the turbulent bubbly fow, the concentration profiles show both a spreading in the horizontal direction and an attenuation of the maximal concentration along the height. At all heights, the profiles remain symmetric since the flow is homogeneous in the horizontal plane. The shape of the distribution is nearly Gaussian, suggesting that we are observing a diffusive process. 

\begin{figure}
	\centering
	\includegraphics[scale=0.5]{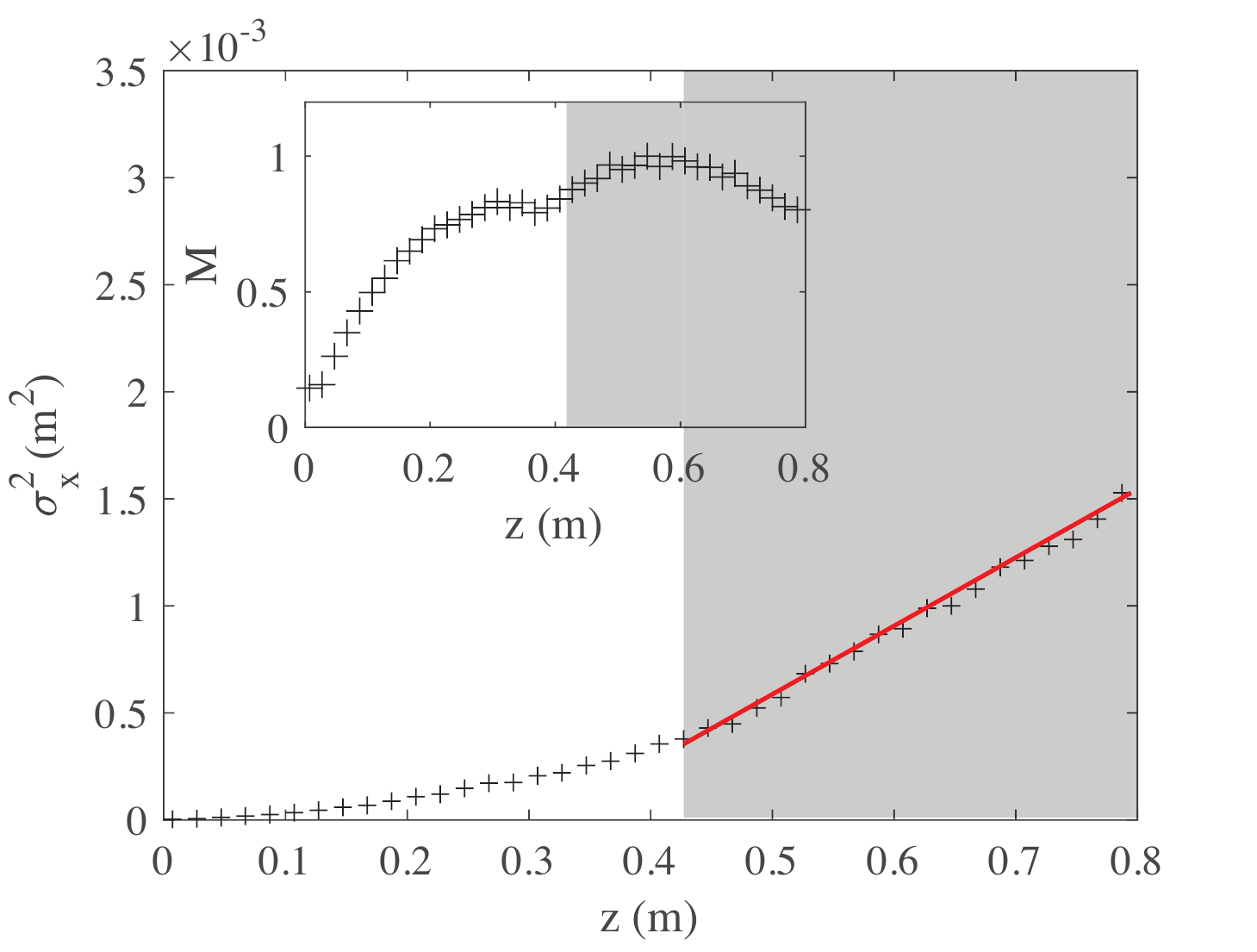}
	\includegraphics[scale=0.5]{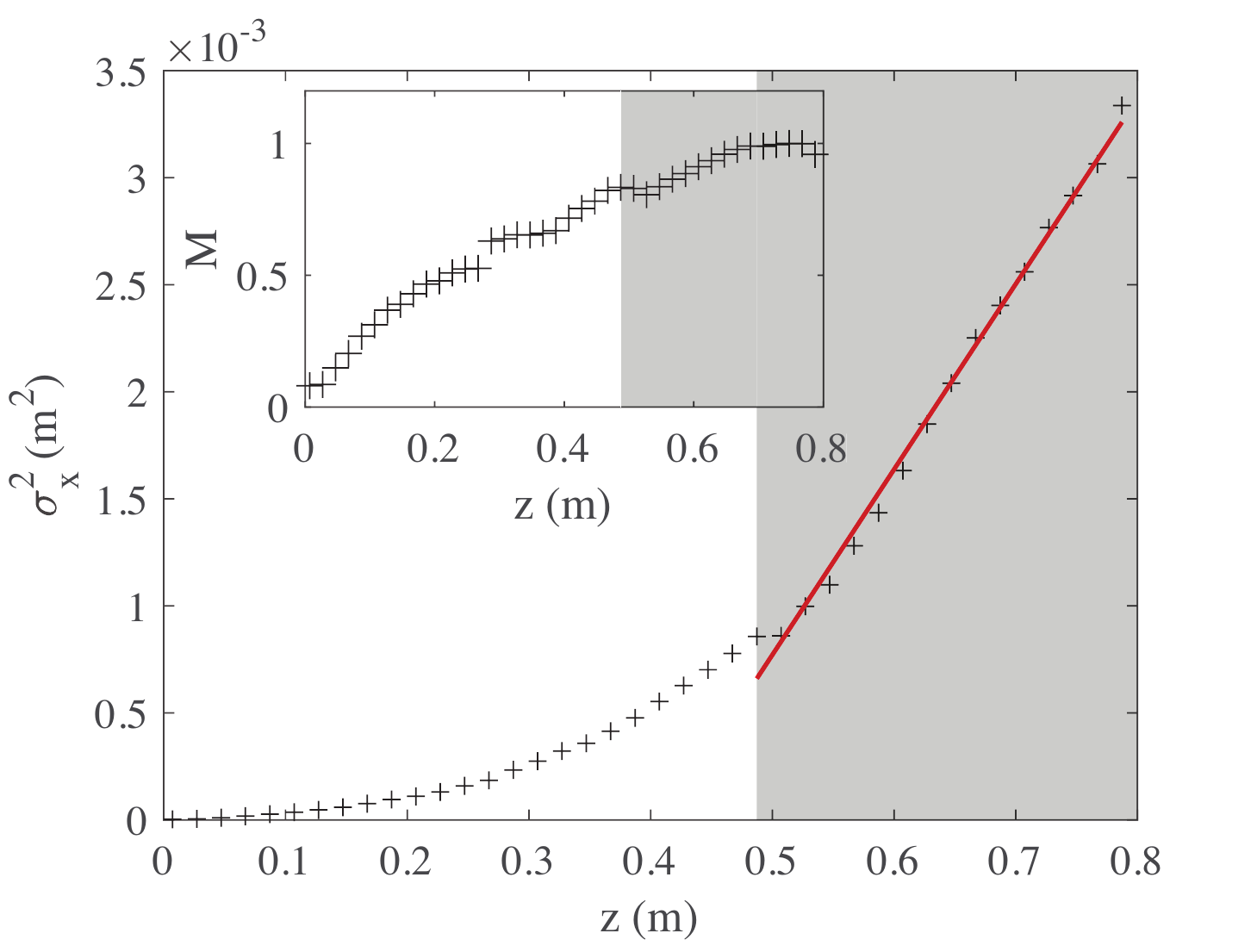}
	\caption{Spatial evolution (in the vertical direction) of the variance of the time-averaged horizontal distribution of the dye for two different bubblance parameters. (a) $b=0.12$ ($u'_0=5.5$cm/s, $\alpha = 0.25\%$) and (b) $b=0.97$ ($u'_0=2.3$cm/s, $\alpha =0.93 \%$). Star symbols: experimental measurement. Red line : Linear fit performed in the range where the mass is nearly conserved, which is highlighted by the gray area. \textit{Insets}: Spatial evolution of the estimated mass flow rate $\dot{M}(z)$ of the dye along the height $z$. Units normalized with the maximum value.}
	\label{fig:std_vs_z}
\end{figure}

The analytical solution of the diffusion equation in a finite three-dimensional medium for a continuous point source, neglecting longitudinal diffusion, can be expressed as follows \citep{socolofskybook}~:
\begin{equation}
c(x,y,z)=\frac{\dot{M}}{ 2\pi U \sigma_x \sigma_y}  \sum \limits^{k=-\infty}_{k=+\infty} exp \left( \frac{-(x+2kw)^2}{2 \sigma_x^2} - \frac{(y+2kw)^2}{2 \sigma_y^2}\right)
\label{eq:eqchpcc}
\end{equation}
where $\dot{M}$ is the mass flow rate of the injected dye, $\sigma_x$ (resp. $\sigma_y$) the standard deviation of the concentration distribution in the $x$-direction (resp. $y$-direction), and $w$ the distance between the dye injector and the walls. It is important to stress that the present measurement technique allows to measure 2D concentration field only, which gives the integrated effect of the depth ($y$) direction. The concentration field viewed by the camera can thus be estimated by integrating equation~(\ref{eq:eqchpcc}) in the $y$-direction between $y=[-w,w]$, leading to~:
\begin{equation}
C(x,z)=\frac{\dot{M}}{ \sqrt{2\pi U} \sigma_x(z)} \sum \limits^{k=-\infty}_{k=+\infty}  exp \left( -\frac{(x+kw)^2}{2 \sigma_x(z)^2} \right)
\label{eq:cc_camera}
\end{equation}


Note that since three sources are sufficient to ensure a good accuracy, the infinite summation has been truncated to $-1\leq k \leq 1$. For each position $z$, the experimental concentration profiles have been fitted to equation~(\ref{eq:cc_camera}) by adjusting independently $\dot{M}$ and $\sigma_x$, using the least-squares method. In figure \ref{fig:concentration_profile}(b), a very good agreement can be observed between the experimental profiles and the fit based on equation~\ref{eq:cc_camera} for all heights $z$. Moreover, the concentration profiles normalised by $\sqrt{2\pi U} \sigma_x(z) / \dot{M}$ collapse when plotted as function of the normalised horizontal position $ x/\sigma_x(z)$. This suggests that the two parameters: $\dot{M}$ and $\sigma_x$ are sufficient to describe the present diffusive process (figure \ref{fig:concentration_profile}(c)). In the following, we will focus on the evolution of the standard deviation $\sigma_x(z)$ of the horizontal concentration distribution along the height $z$, and the measured mass flow rate of dye~$\dot{M}(z)$. \\

Figure \ref{fig:std_vs_z} presents the evolution of the standard deviation $\sigma_x(z)$ of the horizontal concentration distribution and of the mass flow rate $\dot{M}(z)$ along the height $z$ for two cases, $b=0.12$ and $b=0.97$. Far enough from the dye injector located at $z=0$ cm, the standard deviation of the concentration distribution increases linearly with $z$, and the calculated mass flow rate $\dot{M}(z)$ is nearly conserved. This domain is represented in gray in figure~\ref{fig:std_vs_z}. Before this stage is reached, the mass flow rate appears to increase, which is due to highly concentrated dye that absorbes UV neons light close to the dye injector. In this range the dye concentration estimation is not accurate. Measurements are thus restricted to heights where the variation of the measured mass flow rate remains within 15$\%$. In these domains, mixing can be described as a diffusive process since the standard deviation of the concentration profiles $\sigma_x(z)$ evolves linearly with $z$. Using a time to space transformation by means of the liquid mean flow, the effective diffusivity in the horizontal direction can be written as: 
\begin{equation}
D_{xx}=\frac{1}{2} U \frac{d \sigma_x^2}{dz}.
\end{equation}

\begin{figure}[!htbp]
	\centering
	\includegraphics[scale=0.52]{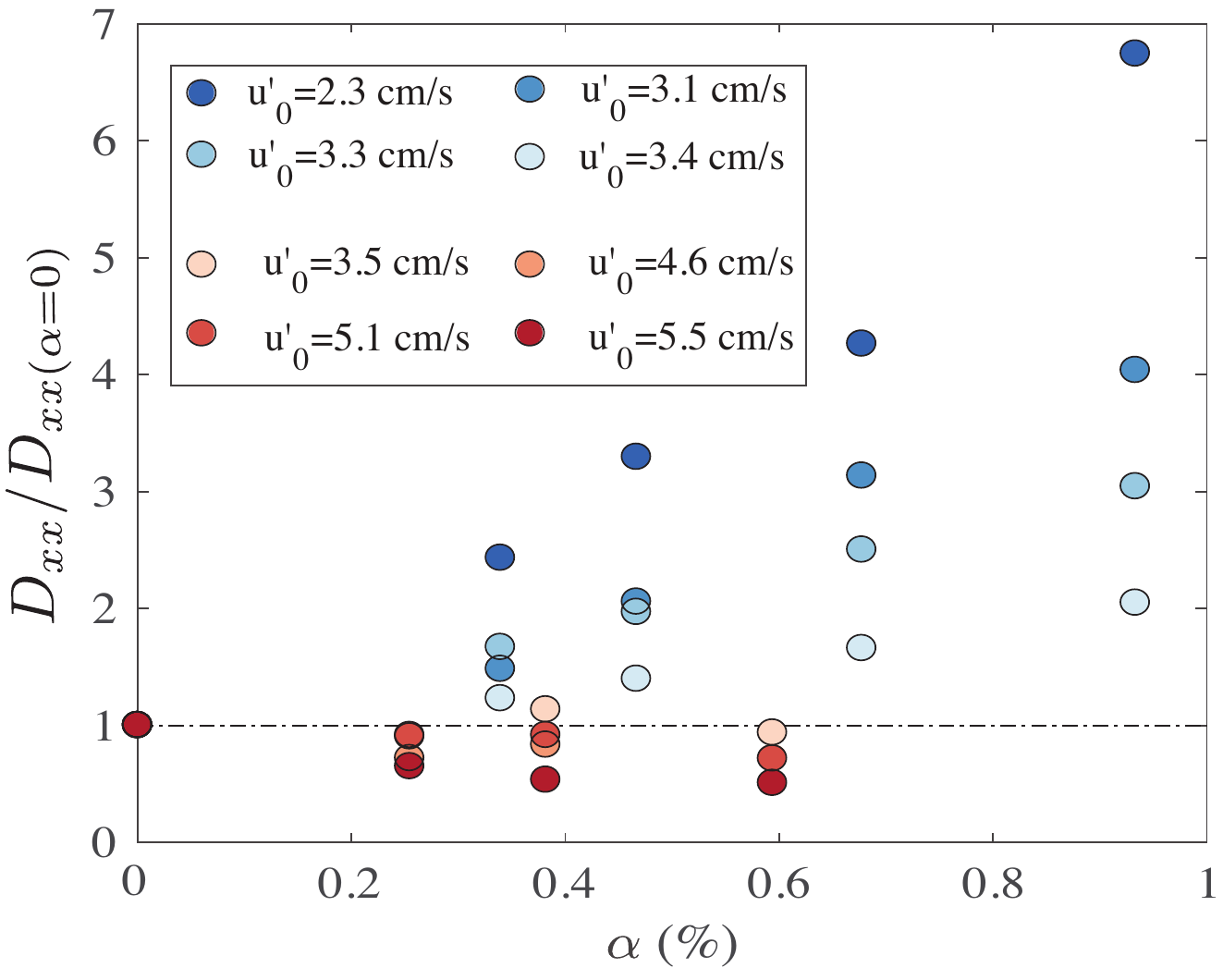}
	\caption{Diffusion coefficient in the horizontal direction normalised by that of the single phase flow as a function of the gas volume fraction $\alpha$ for different turbulence levels $u'_0$.}
	\label{fig:D}
\end{figure}
In this way, we measured the effective diffusion coefficient $D_{xx}$ in the horizontal direction both for single phase and two-phase flows at different turbulence intensities. Hereafter, we will relate the dye dispersion properties to the hydrodynamics in order to get a better understanding of the mixing mechanisms in the present turbulent bubbly flow.

\section{Modelling the effective diffusivity \label{sec:discussion}}

Figure~\ref{fig:D} presents the evolution of the diffusion coefficient normalised by the single phase diffusion coefficient~($\alpha= 0\%$) with increasing the gas volume fraction for different turbulence levels $u_0'$. The diffusion coefficient increases with gas volume fraction at low turbulence levels, while it is nearly constant at high turbulence levels.  The data shows a wide spread with no collapse. To better understand this behaviour we follow the approach of Alm\'eras {\it al.}\cite{JFMalmeras2015}. To begin with, for homogeneous bubbly flows, the velocity fluctuations decorrelate either within the time scale $T_L$ set by the turbulence, or according to the time scale given by the successive passage of bubbles $T_{2b}  =  \frac{d}{\chi^{2/3} V_r \ \alpha}$. 
The bubble passage time scale can be expected to become dominant beyond a critical gas volume fraction $\alpha_c$, such that at $\alpha_c$ we have $T_{2b} \sim T_{L}$. Therefore, the critical gas volume fraction $\alpha_c$ can be calculated as:
\begin{equation}
\alpha_c = \frac{d} {\chi^{2/3}V_r \ T_L }.
\label{eq:alpha_c}
\end{equation}

\begin{figure}
		\centering
		\includegraphics[scale=0.5]{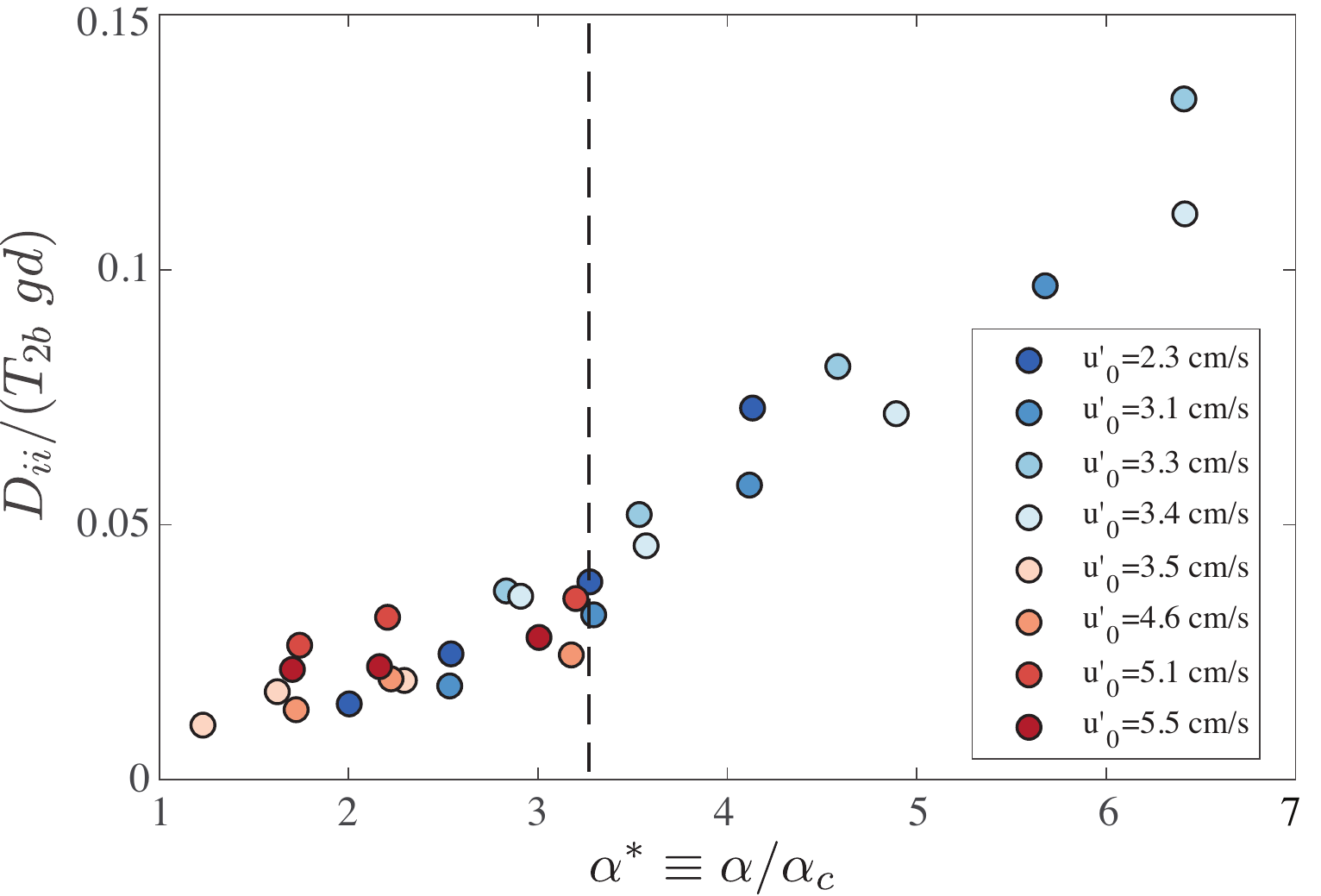}
		\caption{Generalisation of the model of effective diffusivity in bubbly flows. Diffusion coefficient normalised by $T_{2b} gd$ as a function of the gas volume fraction normalised by the critical gas volume fraction $\alpha_c$. Here the normalisation is by the time scale of the bubble passage, $T_{2b}$, and a characteristic velocity scale $g d$, where $g$ is the graviational acceleration and $d$ the bubble diameter. We observe a difference in behavior as the gas volume fraction is increased well above $\alpha_c$.}
		\label{fig:D_allconfig}
	\end{figure}

In figure ~\ref{fig:D_allconfig} we plot the dimensionless diffusion coefficient $D_{xx}/(T_{2b}gd)$ for the bubbly turbulent cases versus the normalized gas volume fraction $\alpha^* \equiv \alpha/\alpha_c$. With this normalisation the effective diffusivities collapse for the various levels of turbulence $u'_0$. Note that the above normalization was also motivated by prior work~\citep{JFMriboux}, which suggests that the liquid agitation in homogeneous bubbly flows should scale as $\sim V_0 \ \alpha^{0.4}$, where $V_0 \propto \sqrt{gd}$ is the rise velocity of an isolated bubble of diameter $d$. In figure~\ref{fig:D_allconfig} we observe two regimes of growth. For large $\alpha^*> 3$, the diffusion coefficient increases at a steeper rate than for smaller $\alpha^*$, indicating the dominance of the bubble passages, which manifests through an increase  in $\alpha^*$. We note that the specific value of the regime transition~(i.e. $\alpha^* \approx 3$) could be influenced by the particulars of the problem such as bubble properties, flow configuration (confined vs three dimensional), and the time scale $T_L$ of the flow.

	\section{Conclusion}
	\label{sec:conc}
	The present paper investigates the mixing of a passive scalar at high Schmidt number in a turbulent bubbly flow. The gas volume fraction $\alpha$ and the velocity fluctuations of the turbulent flow $u'_0$ have been varied in the range $0 \leq \alpha \leq 0.93\%$ and 2.3 $\leq u'_0 \leq 5.6$ cm/s, respectively, resulting in a variation of the bubblance parameter in the range $0 \leq b \leq 1.3$. We continuously injected a low-diffusive dye into the flow, and the fluorescence induced by UV neon tubes was used to measure horizontal concentration profiles at different heights from the dye injector. By approximating the experimental concentration profiles by the solution of the diffusion equation for continuous injection into a finite media, we estimated the effective diffusivities in the horizontal direction for the different operating conditions. 
	
	We find that the growth of the diffusion coefficient can be expressed as a function of the normalized gas volume fraction $\alpha^* \equiv \alpha/\alpha_c$, where $\alpha_c$ is a theoretical critical gas volume fraction. It is remarkable that the theoretically estimated $\alpha_c$ is able to nicely collapse the data. Additionally, we find that the diffusion coefficient growth shows two regimes separated by $\alpha^* \approx 3$. For small gas volume fractions ($\alpha^* < 3$), the diffusion coefficient only weakly increases, while at larger $\alpha^*$, the diffusion coefficient grows more steeply. 
	
	Our study has thus extended the model developed by Alm\'eras {\it et al.}\cite{JFMalmeras2015} to three dimensional turbulent bubbly flows. {\color{black} Additionally, we emphasize the importance of evaluating the critical gas volume fraction $\alpha_c$, which depends on the ratio between the Lagrangian time scale $T_L$ and the bubble passage time scale $T_{2b} \sim \frac{d}{\chi^{2/3}V_r \ \alpha}$, and the operating conditions of the experiment. Thus, we can state that knowledge of the overall liquid agitation and the characteristic time scale of diffusion are key ingredient to predict the mixing in turbulent bubbly flows.}\\
	~~\\

	We thank On-Yu Dung for help with one of the experiments and Gert-Wim Bruggert and Martin Bos for technical support. This work is part of the industrial partnership programme of the Foundation for Fundamental Research on Matter (FOM). The authors also acknowledge the Netherlands Center for Multiscale Catalytic Energy Conversion~(MCEC), STW foundation, European High-performance Infrastructures in Turbulence~(EUHIT), and COST action MP1305 for financial support. Chao~Sun acknowledges the financial support from Natural Science Foundation of China under Grant No. 11672156.

\bibliography{refbiblio}

\end{document}